%% file: LeptonDM-PRL-vfinal.tex
\documentclass[prl,twocolumn,preprintnumbers,nofootinbib,noeprint]{revtex4-2}
\usepackage{amssymb,amsmath,graphicx}
\usepackage[dvipsnames]{xcolor}
\usepackage[colorlinks,allcolors=NavyBlue]{hyperref}
\usepackage{standalone}

\newcommand{\zp}{{Z^\prime}}
\newcommand{\brac}[2]{ \left( \frac{#1}{#2} \right) }

\preprint{FERMILAB-PUB-21-313-T}

\begin{document}

\title{The Simplest and Most Predictive Model of Muon $g-2$ and Thermal Dark Matter}

\author{Ian Holst$^{a,b}$}
\thanks{holst@uchicago.edu, ORCID: https://orcid.org/0000-0003-4256-3680}

\author{Dan Hooper$^{a,b,c}$}
\thanks{dhooper@fnal.gov, ORCID: http://orcid.org/0000-0001-8837-4127}

\author{Gordan Krnjaic$^{a,b,c}$}
\thanks{krnjaicg@fnal.gov, ORCID: http://orcid.org/0000-0001-7420-9577}

\affiliation{$^a$University of Chicago, Department of Astronomy and Astrophysics, Chicago, Illinois 60637, USA}
\affiliation{$^b$University of Chicago, Kavli Institute for Cosmological Physics, Chicago, Illinois 60637, USA}
\affiliation{$^c$Fermi National Accelerator Laboratory, Theoretical Astrophysics Group, Batavia, Illinois 60510, USA}

\date{\today}

\begin{abstract}
The long-standing $4.2 \, \sigma$ muon $g-2$ anomaly may be the result of a new particle species which could also couple to dark matter and mediate its annihilations in the early universe. In models where both muons and dark matter carry equal charges under a $U(1)_{L_\mu-L_\tau}$ gauge symmetry, the corresponding $\zp$ can both resolve the observed $g-2$ anomaly and yield an acceptable dark matter relic abundance, relying on annihilations which take place through the $\zp$ resonance. Once the value of $(g-2)_{\mu}$ and the dark matter abundance are each fixed, there is very little remaining freedom in this model, making it highly predictive. We provide a comprehensive analysis of this scenario, identifying a viable range of dark matter masses between approximately 10 and 100 MeV, which falls entirely within the projected sensitivity of several accelerator-based experiments, including NA62, NA64$\mu$, $M^3$, and DUNE. Furthermore, portions of this mass range predict contributions to $\Delta N_{\rm eff}$ which could ameliorate the tension between early and late time measurements of the Hubble constant, and which could be tested by Stage 4 CMB experiments.
\end{abstract}

\maketitle

\section{Introduction}

Recently, the Fermilab Muon $g-2$ Collaboration presented its first measurement of the muon anomalous magnetic moment~\cite{Albahri_2021}. Their results are consistent with those reported by the previous Brookhaven E821 Collaboration~\cite{Bennett:2006fi}, and the combined world average for $a_\mu \equiv \frac{1}{2} (g-2)_\mu$ now differs from the Standard Model (SM) prediction \cite{Aoyama_2020,Aoyama:2012wk,Aoyama:2019ryr,Czarnecki:2002nt,Gnendiger:2013pva,Davier:2017zfy,Keshavarzi:2018mgv,Colangelo:2018mtw,Hoferichter:2019gzf,Davier:2019can,Keshavarzi:2019abf,Kurz:2014wya,Melnikov:2003xd,Masjuan:2017tvw,Colangelo:2017fiz,Hoferichter:2018kwz,Gerardin:2019vio,Bijnens:2019ghy,Colangelo:2019uex,Blum:2019ugy,Colangelo:2014qya} by
\begin{equation}
\label{amu-exp}
\Delta a_\mu = (251 \pm 59) \times 10^{-11},
\end{equation}
constituting a $4.2 \,\sigma$ discrepancy.\footnote{Although a recent lattice calculation of the SM prediction for $a_\mu$ by the BMW collaboration~\cite{Borsanyi_2021} is closer to the measured value, it is also in tension with the consensus SM prediction, driven by $R$-ratio calculations~\cite{Aoyama_2020}. Future lattice calculations and improved $R$-ratio data will likely be needed to clarify this situation.}

It is well known that MeV-scale particles which couple preferentially to muons can economically resolve the $\Delta a_\mu$ anomaly with muonic couplings of order $g_\mu \sim 10^{-4}$ \cite{Pospelov_2009,Chen_2017,Fayet:2007ua} (for reviews of new physics explanations for the muon $g-2$ anomaly, see Refs.~\cite{Athron:2021iuf,Capdevilla:2021rwo}).\footnote{Dark photons and similar $U(1)$ extensions ({\it e.g.}~gauged $B-L$) featuring appreciable couplings to first generation fermions have been excluded as explanations for $\Delta a_\mu$ \cite{Bauer:2018onh,Ilten_2018}. Similar conclusions hold for light Higgs-mixed scalars and for leptophilic scalars with couplings proportional to their mass~\cite{Krnjaic:2015mbs,Lees_2020}.} It is also well known that such particles can couple to dark matter (DM) and mediate their annihilations to SM particles, potentially leading to the production of an acceptable thermal relic abundance in the early universe \cite{Kamada:2018zxi,Singirala:2021gok,Borah:2021jzu,Biswas:2019twf,Krnjaic:2019rsv,Drees:2018hhs,Okada:2019sbb}. However, the cross section for $t$-channel annihilation,
\begin{equation}
\label{tchan}
\sigma v \sim \frac{ g_\chi^4}{ m_\chi^2 } \sim 10^{-25} \, {\rm cm}^3 {\rm s}^{-1} \brac{g_\chi}{10^{-2}}^4
\brac{\rm GeV}{m_\chi}^2,
\end{equation}
can only be sufficiently large for thermal production if $g_\chi \gg g_\mu$.
A similar conclusion holds for $s$-channel annihilation, for which the cross section scales as $ \propto g_\chi^2 g_\mu^2$, thus requiring $g_\chi$ to be even larger~\cite{Kahn:2018cqs,Blinov:2020epi,Krnjaic:2019rsv}. While scenarios with $g_\chi \gg g_\mu$ are not ruled out, this freedom reduces the predictive power of such models and imposes inelegant requirements on the gauge groups involved, posing significant challenges for their embedding in non-Abelian theories.

An exception to this lore was identified in Ref.~\cite{Foldenauer:2018zrz}, which exploits resonant annihilation in the context of a spontaneously broken $U(1)_{{L_\mu - L_\tau}}$ gauge group under which both the muon and the DM have comparable couplings to the corresponding gauge boson, $\zp$. Motivated by the latest measurement of muon $g-2$, we comprehensively analyze this scenario, identifying new parameter space that can both explain $\Delta a_\mu$ and yield an acceptable DM abundance for mass ratios of $m_{\zp}/m_{\chi } \sim 2- 3$. Once the gauge coupling is fixed by the measured value of $\Delta a_\mu$ and the $m_\chi/m_{\zp}$ ratio is fixed by the measured DM abundance, we are left with a one-parameter family of models, making this scenario highly predictive. A combination of future laboratory measurements and cosmological observations will conclusively discover or falsify this scenario.

\section{Model Overview}

We begin by reviewing the basic features of gauged $L_\mu - L_\tau$ extensions of the SM, as introduced in Refs.~\cite{He:1990pn,He:1991qd}. The Lagrangian for this scenario includes
\begin{eqnarray} \label{eq:lag1}
{\cal L}_{\rm int} = g^\prime Z^\prime_\mu ( J^{\mu}_{\rm \small SM} + J_{\rm DM}^\mu) , ~~
\end{eqnarray}
where $g^\prime$ is a universal gauge coupling and $\zp$ is the $L_\mu - L_\tau$ gauge boson. We assume that this gauge symmetry is spontaneously broken in the infrared, but that the states responsible for this breaking are sufficiently decoupled that their effects are negligible at low energies. The SM current in Eq.~\ref{eq:lag1} is given by
\begin{eqnarray}
\label{eq:Jmutau}
J^{\nu}_{\rm SM} =  \bar \mu\gamma^\nu \mu +  \bar \nu_\mu \gamma^\nu P_L \nu_{\mu}  -
                \bar \tau\gamma^\nu \tau -  \bar \nu_\tau \gamma^\nu P_L \nu_{\tau} ,
\end{eqnarray}
where $P_L $ is the left chiral projector. We take the DM, $\chi$, to be a Dirac fermion, with $J_{\rm DM}^\mu \equiv \bar \chi \gamma^\mu\chi$ (for the case of DM in the form of a complex scalar, see the Supplemental Material).\footnote{Throughout this study, we will assume that the DM carries one unit of $L_{\mu}-L_{\tau}$ charge. If another charge assignment is selected, the effective coupling of the $\zp$ to the DM will be shifted by this quantity.}

\begin{figure}[t]
  \hspace{-0.3cm}
  \includegraphics[width=8.5cm]{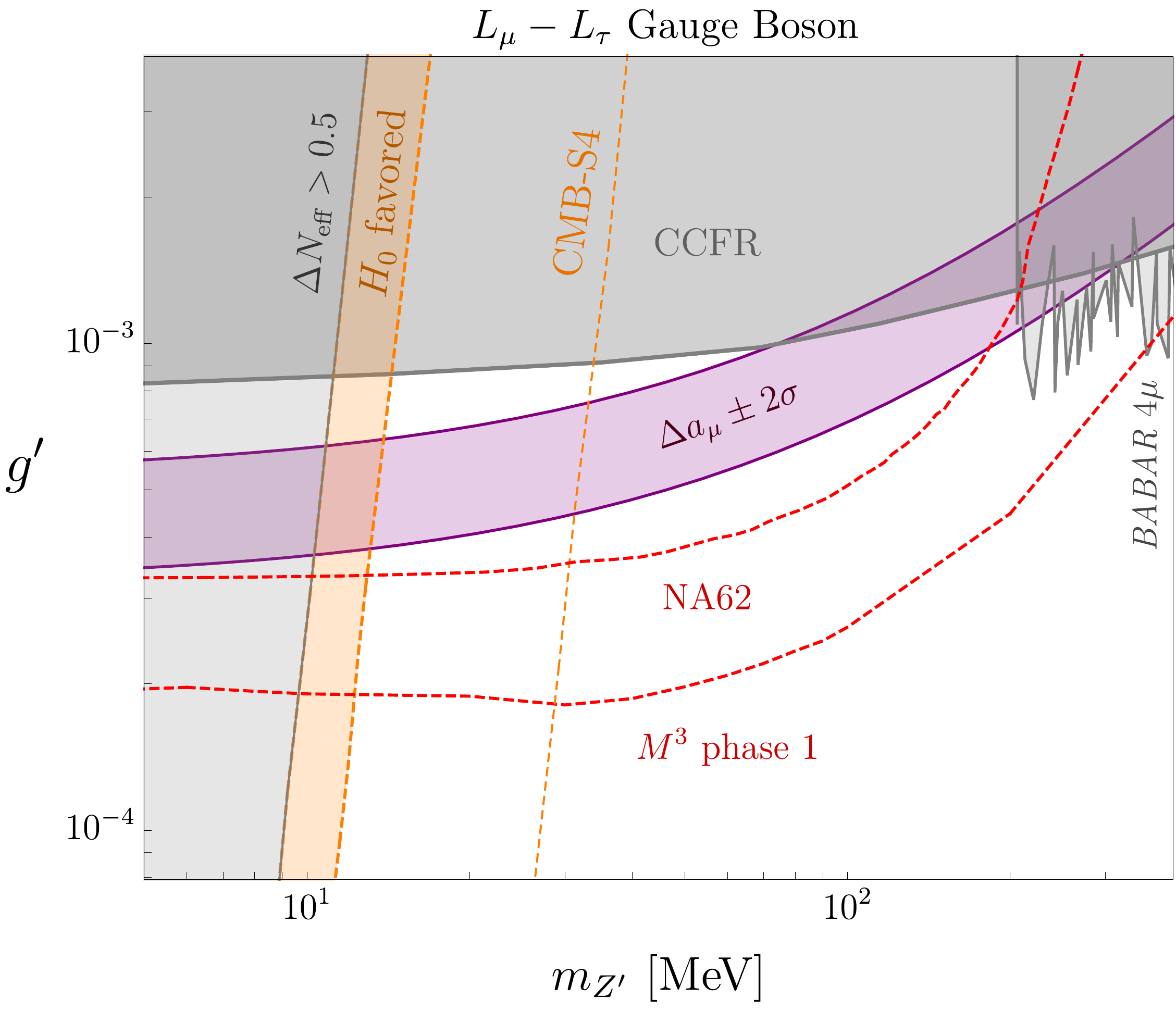}
  \caption{The purple band represents the regions of parameter space in which an $L_\mu - L_\tau$ gauge boson can resolve the muon $g-2$ anomaly (within $\pm 2 \sigma$). Inside the nearly vertical orange band, this model can also relax the $H_0$ tension by contributing $\Delta N_{\rm eff} \sim 0.2-0.5$ \cite{Escudero:2019gzq}; the mild slope arises from delayed neutrino decoupling via loop induced $\zp$-$\gamma$ kinetic mixing. The gray shaded regions represent laboratory constraints from CCFR~\cite{PhysRevLett.66.3117,Altmannshofer:2014pba}, BABAR~\cite{Lees_2016}, and the constraint of $\Delta N_{\rm eff} > 0.5$ from measurements of the CMB and the primordial light element abundances~\cite{Escudero:2019gzq}. Note that here we only show the contribution to $\Delta N_{\rm eff}$ from from the $\zp$, without any additional contribution from dark matter annihilation. Also shown are projected sensitivity of the NA62 experiment~\cite{Krnjaic:2019rsv}, and phase 1 of the $M^3$ muon missing momentum experiment~\cite{Kahn:2018cqs}.}
  \label{fig:vectorfig}
  \vspace{0cm}
\end{figure}

The partial width for $\zp \to f\bar{f}$ decays is given by
\begin{eqnarray}
\label{gamma}
\Gamma_{\zp \to f\bar f} =  \frac{k_f {g^\prime}^2 m_{\zp}}{12\pi } \left( 1 + \frac{2m^2_f}{m^2_{\zp}} \right)     \sqrt{1 - \frac{4 m_f^2}{m^2_{\zp} }}~~,
\end{eqnarray}
where $k_f = 1$ for $f = \mu,\tau,\chi$ and $k_f = 1/2$ for $f =\nu_\mu, \nu_\tau$. Note that
the interactions in Eq.~\ref{eq:Jmutau} also lead to an irreducible contribution to the kinetic mixing of the $\zp$ with the photon through $\mu$ and $\tau$ loops, of order $\varepsilon \sim g^{\prime}/70$~\cite{Escudero:2019gzq}.
In Fig.~\ref{fig:vectorfig}, we show the viable parameter space for this $\zp$.

The leading contribution to $a_\mu$ in this model arises at one loop where
\begin{eqnarray}
\label{delta-amu}
\Delta a_{\mu} = \frac{{g^\prime}^2}{4\pi^2} \! \int_0^1 \! dx \frac{ m^2_\mu x (1-x)^2}{ m_\mu^2 (1-x)^2 +  m_{\zp}^2 \, x},
\end{eqnarray}
which yields $\Delta a_\mu \approx 251\times 10^{-11} \times ({g^\prime / 4.5 \times 10^{-4}})^2$  in the $m_{\zp} \ll m_\mu$ limit~\cite{Pospelov_2009}. As we will discuss below, for couplings that resolve $\Delta a_\mu$,
this scenario is viable for $m_{\zp} \sim$\,$10-200$ MeV. For the remainder of this {\it Letter}, we will restrict our attention to this range of masses.

\section{Cosmology}

To resolve the $\Delta a_\mu$ anomaly, the combination of Eqs.~\ref{amu-exp} and \ref{delta-amu} implies $g^\prime \gtrsim 10^{-4}$ for all viable values of $m_{\zp}$, as shown in Fig.~\ref{fig:vectorfig}. For couplings of this size, $\zp$ and $\chi$ are both easily maintained in chemical equilibrium with the SM in the early universe, so achieving the measured DM density requires an annihilation cross section of order $\sigma v\sim 5 \times 10^{-26}$ cm$^3$ s$^{-1}$ (for $m_\chi <$ GeV)~\cite{PhysRevD.86.023506}.

From Eq.~\ref{tchan}, it is clear that this cross section cannot be achieved with $t$-channel $\chi \bar \chi \to \zp\zp$ annihilation when the natural coupling relation is enforced, $g^\prime = g_\mu = g_\chi$, and the DM mass is large enough to be consistent with constraints from Big Bang nucleosynthesis~\cite{Boehm:2013jpa,PhysRevD.91.083505}. For appropriate values of $m_{\chi}/m_{\zp}$, however, it is possible to achieve an acceptable relic abundance through $s$-channel resonant annihilation.

\begin{figure*}[t]
  \hspace{-0.3cm}
  \includegraphics[width=8.35cm]{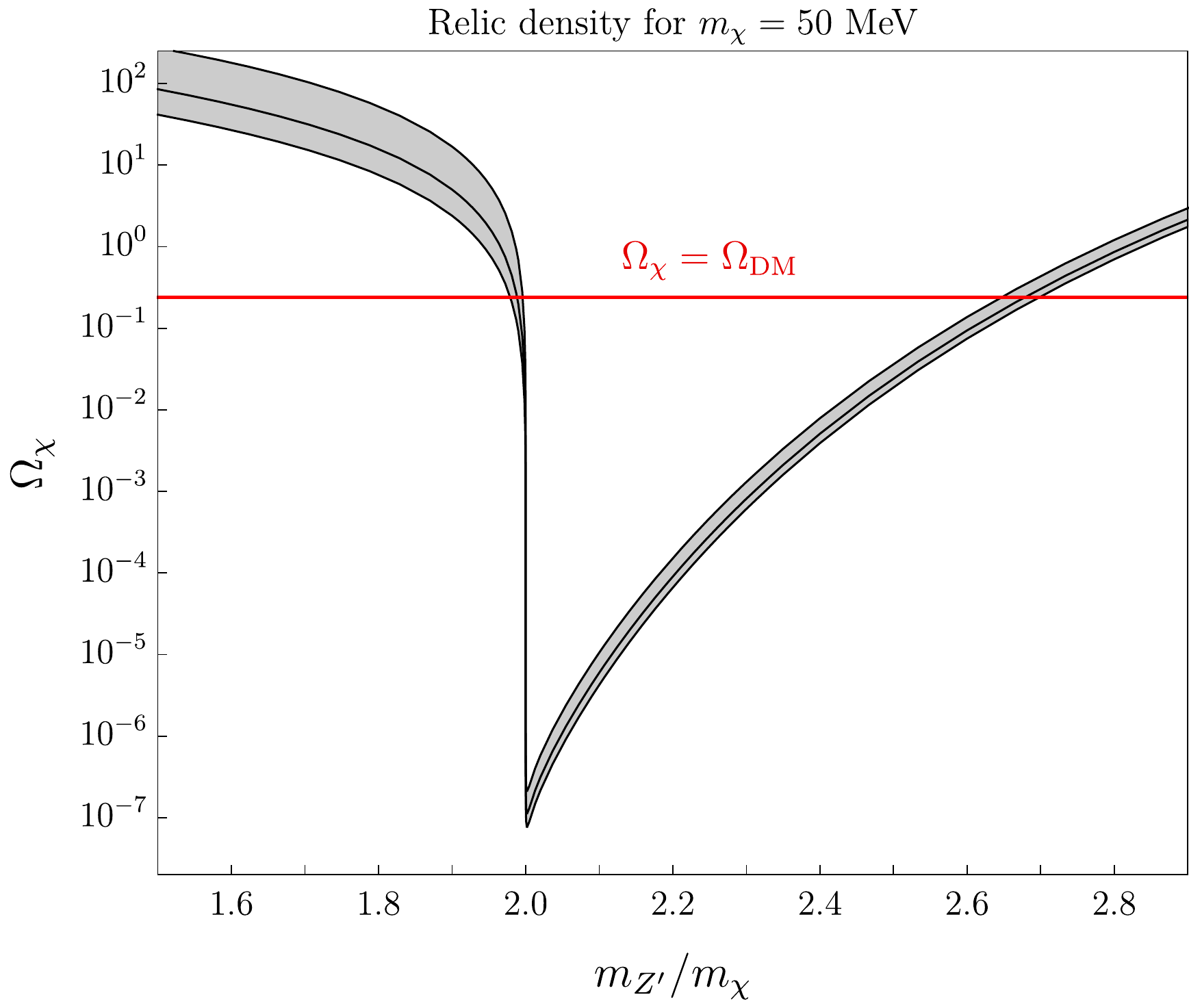} ~~
  \includegraphics[width=8.6cm]{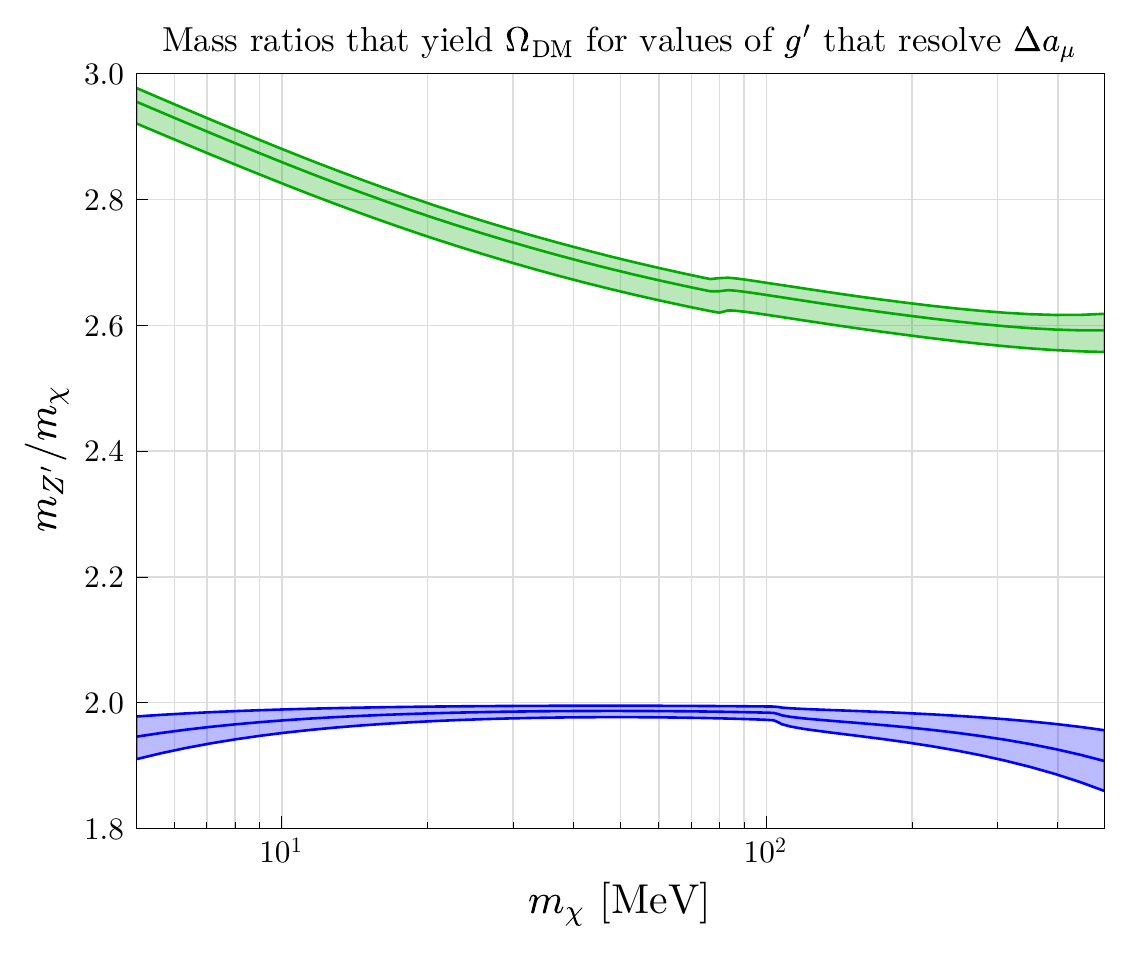}
  \caption{{\bf Left:} The dark matter relic abundance as a function of $m_{\zp}/m_{\chi}$, for $m_{\chi}=50\, {\rm MeV}$, and for values of $g'$ selected to resolve the muon $g-2$ anomaly (the thickness of the band reflects the $\pm 2\sigma$ uncertainty in the measurement of $a_{\mu}$). For each choice of $m_\chi$, there are two values of $m_{\zp}$ that can achieve the observed DM abundance indicated by the horizontal red line. {\bf Right:} $m_{\zp}/m_\chi$ ratios that yield the observed DM density for the couplings that resolve $\Delta a_\mu$ anomaly to within $\pm 2\sigma$ at each value of $m_{\zp}$; the two bands represent the two solutions for each $m_\chi$ as shown in the left panel. Data is available at Ref.~\cite{zenodo_dataset_2021}.}
  \label{fig:omega-fig}
  \vspace{0cm}
\end{figure*}

%\subsection{Dark Matter Freeze Out}

In the very early universe, the total number density of DM particles, $n \equiv n_\chi + n_{\bar \chi}$, is governed by the  Boltzmann equation:
\begin{eqnarray}
\label{boltzmann}
\frac{ dn}{dt} + 3 H n = - \frac{1}{2}\langle \sigma v\rangle ( n^2- n^2_{\rm eq}),
\end{eqnarray}
where $H$ is the Hubble rate and  $n_{\rm eq}$ is the equilibrium number density. The thermally averaged annihilation cross section is~\cite{GG91}
\begin{eqnarray}
\label{sigmavTA}
\langle \sigma v\rangle =
\frac{1}{8 m_\chi^4 T  K^2_2\left( \frac{m_\chi}{T} \right)}
\int_{4m_\chi^2}^\infty \!\! ds \, \sigma \sqrt{s} \, (s \! - \! 4m_\chi^2)
 K_1 \! \left( \frac{\sqrt{s}}{T} \right)  \!,~~~~
\end{eqnarray}
where $K_n$ is a modified Bessel function of the second kind.
For $m_{\zp} > m_\chi$, DM annihilation proceeds mainly\footnote{While annihilations through the ``forbidden'' channel, $\chi \bar \chi \to \zp \zp$, may also be possible~\cite{Griest:1990kh,D_Agnolo_2015}, this is subdominant in the parameter space close to the resonance region.} through $s$-channel processes, $\chi\bar \chi \to \bar f f$, with a cross section that is given by
\begin{eqnarray}
\sigma(s) = \sum_f
\frac{k_f {g^\prime}^4}{12\pi s}
%\sqrt{\frac{1-4m_f^2/s}{1-4m^2_{\chi}/s}}
\frac{\beta_f}{\beta_\chi}
\left[\frac{
 (s+ 2 m_\chi^2) (s + 2m_f^2)
}{ (s- m^2_{\zp})^2 + m_{\zp}^2 \Gamma^2_{\zp}} \right],~~~~
\end{eqnarray}
where $s$ is the Mandelstam variable, $\beta_i = \sqrt{1-4m_i^2/s}$, and
$\Gamma_{\zp} = \Gamma_{\zp \to \chi \bar \chi} + \sum_f \Gamma_{\zp \to f\bar f}$ with the sums over $f=\mu, \tau, \nu_{\mu,\tau}.$

\begin{figure}[t]
  \hspace{-0.3cm}
  \includegraphics[width=8.4cm]{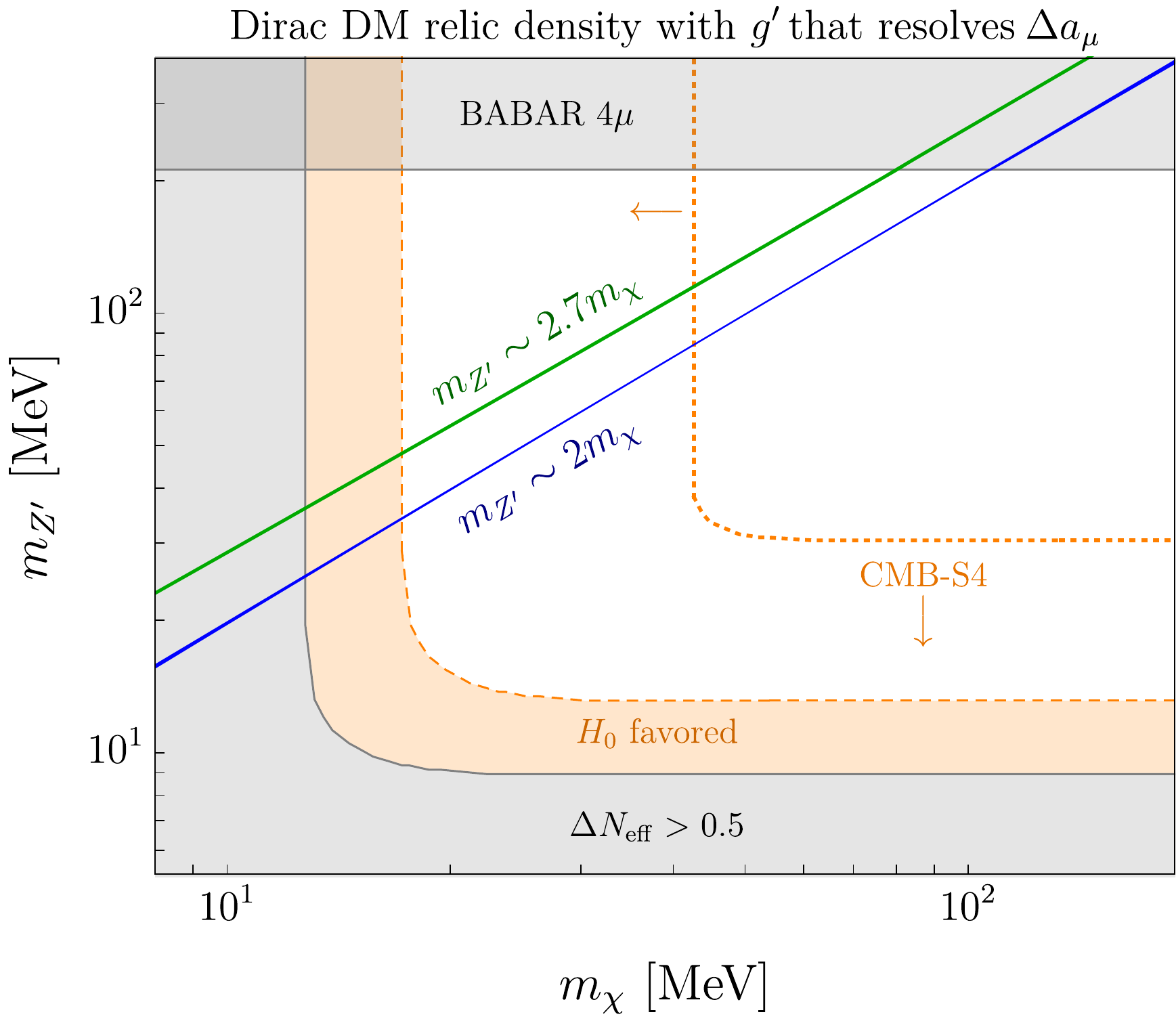}
  \caption{In the green and blue regions, the thermal relic abundance is equal to measured density of dark matter, $\Omega_{\chi} h^2 \simeq 0.12$ \cite{Planck2018}, for values of $g^\prime$ that can resolve the muon $g-2$ anomaly to within $\pm 2 \sigma$. The gray shaded regions are excluded by BABAR \cite{Lees_2016} and by $\Delta N_{\rm eff} >0.5$~\cite{Escudero:2019gzq}. The orange shaded region can ameliorate the $H_0$ tension by contributing $0.2 <\Delta N_{\rm eff} < 0.5$, and the parameter space under and to the left of the dotted orange curve predicts $\Delta N_{\rm eff} > 0.02$, which is projected to be testable with Stage 4 CMB experiments~\cite{CMB-S4:2016ple}. Data for the green and blue regions is available at Ref.~\cite{zenodo_dataset_2021}.}
  \label{fig:mainfig}
  \vspace{0cm}
\end{figure}

In the left panel of Fig.~\ref{fig:omega-fig}, we determine the DM relic abundance $\Omega_\chi$ by numerically\footnote{Our numerical results were obtained using FeynCalc~\cite{MERTIG1991345,SHTABOVENKO2016432,SHTABOVENKO2020107478} and Astropy~\cite{astropy:2013,astropy:2018}.} solving Eq.~\ref{boltzmann} as a function of $m_{\zp}/m_\chi$ and fixing the coupling to obtain the measured value of $(g-2)_{\mu}$ for a given value of $m_\zp$. In the right panel we show the $m_{\zp}/m_\chi$ ratios that yield the observed density as a function of $m_\chi$, with couplings chosen to resolve $\Delta a_\mu$ to within $\pm 2 \sigma$.
In Fig.~\ref{fig:mainfig}, we show the regions in the $m_{\chi}-m_{\zp}$ plane that lead to an acceptable relic abundance, again fixing the coupling according to $(g-2)_{\mu}$. From these figures, we see that the measured density of DM can be obtained for either $m_{\zp}/m_{\chi} \approx 1.9 - 2.0$ or $m_{\zp}/m_{\chi} \approx 2.6-3.0$, which are shown in blue and green in Fig.~\ref{fig:mainfig}, respectively.
While the first band may be considered fine-tuned, we regard the other band as a prediction. Note that our results differ from the mass ratio indicated in Fig.~1 of Ref.~\cite{Foldenauer:2018zrz}.

%\subsection{BBN and $H_0$}

In the mass range of interest, DM annihilations and $\zp$ decays can each heat the SM radiation bath after the neutrinos have decoupled, thereby increasing the effective number of neutrino species,
\begin{eqnarray}
\Delta N_{\rm eff} \equiv \frac{8}{7} \brac{11}{4}^{4/3} \frac{\delta \rho_\nu}{\rho_\gamma}
 \biggr |_{T = T_{\rm dec}}.
\end{eqnarray}
The extra energy density, $\delta\rho_\nu$, receives contributions from both $\chi \bar \chi \to \bar \nu \nu$ and $\zp \to \bar \nu \nu$, which take place after neutrino decoupling\footnote{
Note that $\chi$ and $\zp$ entropy transfers that contribute to $\Delta N_{\rm eff}$  occur while the DM is still in chemical equilibrium
and freeze out occurs below $T_{\rm dec}$, though the
DM number density is already Boltzmann suppressed at $T_{\rm dec}$. Thus, by energy conservation, the total
energy density in Eq \ref{delta-rho} must be transferred to neutrinos; since the DM is particle/antiparticle
symmetric and in equilibrium at $T_{\rm dec}$, there is no chemical potential in Eq \ref{delta-rho}.
}
at $T_{\rm dec} \approx $ 2 MeV, leading to
\begin{eqnarray}
\label{delta-rho}
\delta \rho_\nu \approx  \int \! \frac{d^3\vec p_\chi}{(2\pi)^3}
 \frac{4 E_{\chi}}{e^{E_{\chi}/T_{\rm dec}} + 1} +
 \int  \frac{d^3\vec p_{\zp}}{(2\pi)^3}   \frac{3E_{\zp}}{e^{E_{\zp}/T_{\rm dec}} - 1}~.~~~~~~
\end{eqnarray}
Taken at face value, measurements of the CMB and baryon acoustic oscillations constrain $\Delta N_{\rm eff} \lesssim 0.3$~\cite{Planck2018}. That being said, it has been shown that a positive value of $\Delta N_{\rm eff}$ could help to reduce the long-standing tension between local and cosmological determinations of the Hubble constant, $H_0$~\cite{di_valentino_2021}. When local measurements of $H_0$ are included in the fit, the data prefer $\Delta N_{\rm eff} \sim 0.1-0.4$~\cite{Planck2018}. With this in mind, we require that $\Delta N_{\rm eff} \lesssim 0.5$, and consider any parameter space in which $\Delta N_{\rm eff} = 0.2 - 0.5$ to be potentially capable of reducing the Hubble tension. In Fig.~\ref{fig:mainfig}, we show isocontours of $\Delta N_{\rm eff}$ whose shapes are computed using the approximate expression in Eq.~\ref{delta-rho} and scaled to match the more precise results given in Ref.~\cite{Escudero:2019gzq}.\footnote{In the $m_\chi \gg m_{\zp}$ limit, only $\zp$ contributes to $N_{\rm eff}$, matching the scenario explored in Ref.~\cite{Escudero:2019gzq}. In the opposite limit, only $\chi$ contributes, so $N_{\rm eff}$ can be extracted by rescaling the spin degrees-of-freedom between $\chi$ and $\zp$ in the limit of Maxwell-Boltzmann statistics.}

%\subsection{CMB}

Well after freeze out, the annihilation of DM particles could potentially alter the CMB's temperature anisotropies by injecting visible energy into the radiation bath during and after the era of recombination. For Dirac fermion DM, the Planck collaboration places the following constraint on such annihilations~\cite{Planck2018}:
\begin{eqnarray}
p_{\rm ann} \equiv \frac{1}{2}f_{\rm eff} \frac{\langle \sigma v \rangle}{m_\chi} < 3.5 \times 10^{-28} \,  \frac{ \text{cm}^3 }{   \rm GeV\, s} ~,
\end{eqnarray}
where $\langle \sigma v \rangle$ is the thermally averaged annihilation cross section at recombination, and the quantity $f_{\rm eff}$ parametrizes the efficiency of energy transfer to SM particles.

\begin{figure}[t]
  \hspace{-0.3cm}
  \includegraphics[width=8.5cm]{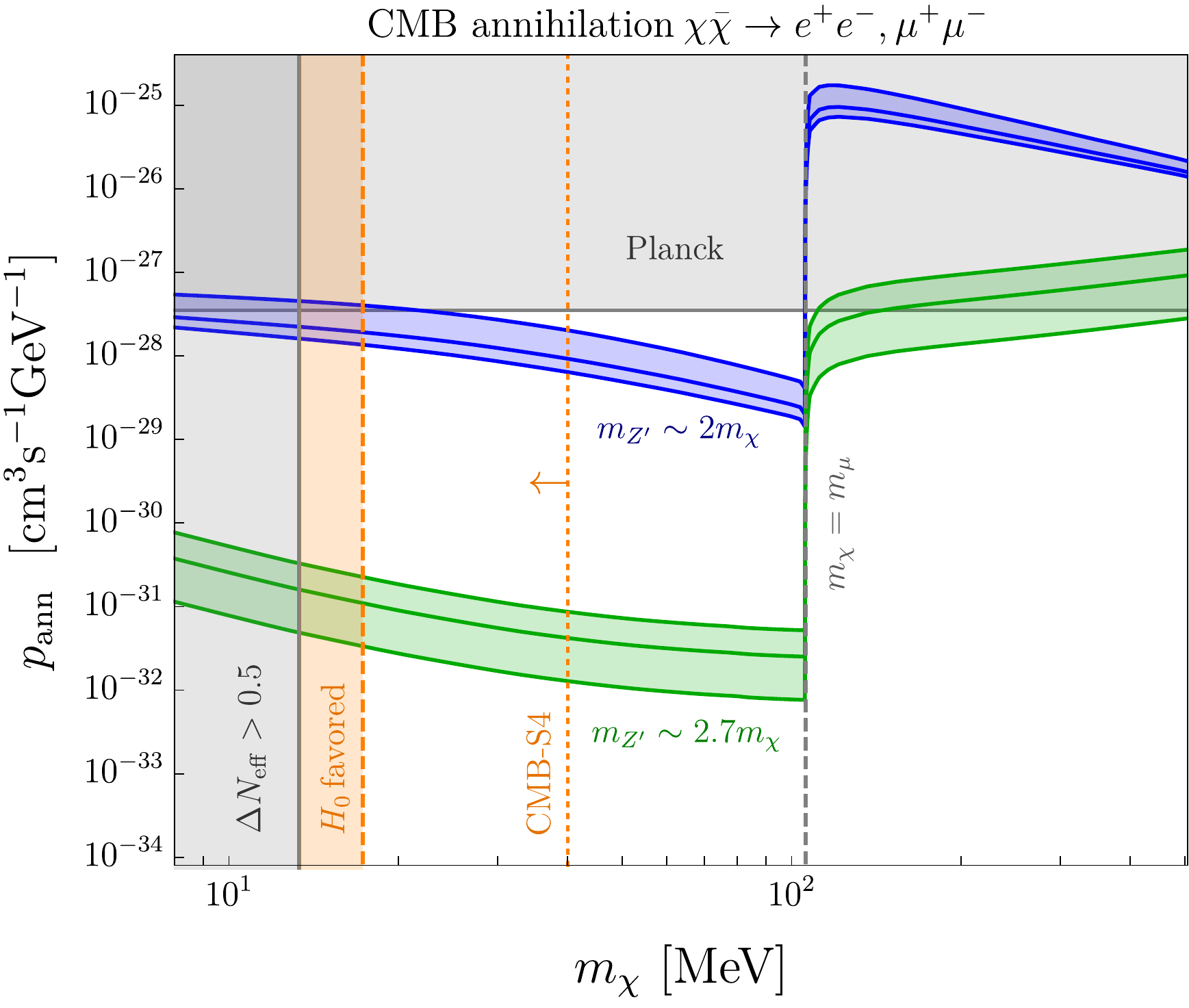}
  \caption{Energy injection from Dirac dark matter annihilating to $e^+e^-$ (through loop-induced kinetic mixing) and/or $\mu^+ \mu^-$, during and after recombination. Following Fig.~\ref{fig:mainfig}, the blue and green regions correspond to values of $m_{\zp}/m_{\chi}$ which produce the measured density of dark matter for values of $g^\prime$ that can resolve the muon $g-2$ anomaly to within $\pm 2 \sigma$.}
  \label{fig:cmb-fig}
  \vspace{0cm}
\end{figure}

In our model, the only channels with non-negligible values of $f_{\rm eff}$ are those of $\chi \bar \chi \rightarrow \mu^+ \mu^-$ (for which $f_{\rm eff} \sim 0.2$) and $\chi \bar \chi \rightarrow e^+ e^-$ ($f_{\rm eff} \sim 1$) \cite{Slatyer_2016}. As we will discuss below, the parameter space of this model in which $m_{\chi} > m_{\mu}$ is almost entirely ruled out by laboratory constraints.\footnote{While there are limits on DM annihilation to muons in similar models from the AMS-02 experiment \cite{zu2021}, that annihilation channel is already excluded in our model. AMS-02 is also not sensitive to the range of $m_\chi$ that we consider.} Because the $\zp$ does not couple directly to electrons, $\chi \bar \chi \rightarrow e^+ e^-$ occurs only through kinetic mixing, with a low-velocity cross section given by
\begin{eqnarray}
\label{sigmav-cmb}
\hspace{-0.051cm}
\langle  \sigma v \rangle_{\chi\bar\chi \to e^+e^-} \simeq \frac{
(  \varepsilon e g^{\prime} )^2 \left(m_e^2 + 2m_\chi^2\right) \sqrt{1-m_e^2/m_\chi^2} }{ 2\pi \left[  \left( m_{\zp}^2 - 4m_\chi^2  \right)^2 + m_{\zp}^2 \Gamma_{\zp}^2  \right]  }.~~~~~
\end{eqnarray}
In Fig.~\ref{fig:cmb-fig}, we show the predicted value of $p_{\rm ann}$ for $\varepsilon = g^{\prime}/70$ and values of $m_{\zp}/m_\chi$ that yield the measured density of DM. We note that while Planck's sensitivity to $p_{\rm ann}$ is approaching the limit of cosmic variance, future CMB experiments are expected to modestly improve upon these constraints~\cite{Green:2018pmd}.

\section{Laboratory Tests}

%\subsection{Neutrino Tridents}

The $\zp$ in this model can mediate inelastic neutrino-nucleus scattering, leading to dimuon production, $\nu_\mu N \to \nu_\mu N \mu^+\mu^-$. The CCFR experiment has measured trident production for muon neutrinos and places a limit of $m_{\zp} \lesssim 300$ MeV for couplings that can resolve the muon $g-2$ anomaly~\cite{PhysRevLett.66.3117,Altmannshofer:2014pba}.

%\subsection{$B$-Factories}

The properties of the $\zp$ can also be constrained by $B$ factories, through the $e^+e^- \to 4 \mu$ channel. For couplings that resolve the muon $g-2$ anomaly, the BABAR experiment excludes 212 MeV $< m_{\zp} <$ 10 GeV \cite{Lees_2016}. Although our scenario includes an additional $\zp \to \chi \bar \chi$ decay channel throughout our mass range of interest, the phase space suppression in the parameter space of interest here renders this width negligible, allowing us to safely apply the BABAR limit.

%\subsection{Direct Detection}

Since the DM does not couple directly to quarks or electrons in this model, direct detection experiments can be expected to only weakly constrain this class of scenarios. The largest contribution to the elastic scattering of DM with nuclei occurs in this model through the kinetic mixing of the $Z'$ with the photon. For $\varepsilon = g^{\prime}/70$, the cross section for DM scattering with protons is
\begin{eqnarray}
\! \sigma_e = \frac{(\varepsilon e g^\prime)^2 m_{\chi}^2}{\pi m_{\zp}^4}
\approx  \! 10^{-48} \, {\rm cm}^2 \! \brac{g^\prime}{10^{-4}}^4 \!\! \brac{m_\chi}{100 \, \rm MeV}^2 \!\!,~~~~~~
\end{eqnarray}
where we have used $m_\chi = 2 m_{\zp}$. The cross section with electrons is further suppressed by a factor of $(m_e/m_\chi)^2 \ll 1$. Existing direct detection experiments are thus not sensitive to the scenario considered here.

\section{Summary}

In this {\it Letter}, we have identified a simple and highly predictive model that can simultaneously resolve the muon $g-2$ anomaly and yield an acceptable dark matter relic abundance. This model introduces a leptophilic $L_\mu -L_\tau$ gauge boson whose coupling to DM and SM particles is fixed by the measured value of $a_{\mu}$. Furthermore, to obtain an acceptable DM relic abundance through resonant annihilations in the early universe, we must fix either $m_{\zp}/m_{\chi} \approx 1.9-2.0$ or $m_{\zp}/m_{\chi} \approx 2.6-3.0$. The viable range of parameter space features DM with masses in the range of $m_\chi \approx 13 - 100 \, {\rm MeV}$. There is no need for ad hoc hierarchical charge assignments in this model, and it is naturally safe from CMB constraints.

A number of accelerator-based experiments are poised to decisively cover the relevant parameter space in this scenario. In particular, the NA62 experiment is sensitive to invisibly decaying muon-philic bosons (through $K\to \mu \nu \zp$), and may be able to cover the entire parameter space favored by the measured value of $\Delta a_\mu$ if systematic errors can be adequately controlled~\cite{Krnjaic:2019rsv}. Dedicated muon beam efforts, including the CERN NA64$\mu$~\cite{Gninenko_2020} and Fermilab $M^3$~\cite{Kahn:2018cqs} experiments, are projected to cover all of the remaining parameter space for any light, invisibly decaying particles that could potentially be responsible for $\Delta a_\mu$. Finally, the upcoming DUNE experiment~\cite{abi2020deep,Ballett_2019} will improve upon existing bounds on muon trident production, covering the remaining $L_\mu - L_\tau$ parameter space that can address the muon $g-2$ anomaly. Our scenario predicts that {\it all} of these experiments will see a signal; a null result at any one of them would decisively falsify this model. Measurements at future fixed-target experiments could also potentially measure $m_{\zp}$~\cite{Blinov:2020epi}, allowing us (through the relic abundance) to infer the required mass of the DM candidate.

However, while laboratory searches can probe the existence of $\zp$ and measure $g^\prime$ to verify this solution to $\Delta a_\mu$, they cannot directly establish its possible connection to the DM of our universe. To test this aspect of our model, it will be necessary to precisely measure both $\Delta N_{\rm eff}$ and $p_{\rm ann}$ with future CMB experiments. From Fig.~\ref{fig:mainfig}, along the viable diagonal bands there is a one-to-one correspondence between $\Delta N_{\rm eff}$ and $m_\chi$, so combined with a laboratory measurement of $m_{\zp}$, an observation of $\Delta N_{\rm eff}$ fixes the $m_{\zp}/m_\chi$ ratio. Since the value of $g^\prime$ required to resolve $\Delta a_\mu$ is also fixed by $m_{\zp}$, there is no additional freedom to vary the annihilation cross sections in this model. Thus, the predicted relic density from thermal freeze out can be verified to agree with the measured value of $\Omega_\text{DM}$ and, once the mass ratio is determined, there is a unique prediction for $p_{\rm ann}$, as shown in Fig.~\ref{fig:cmb-fig}.

% \enlargethispage{\baselineskip}

\bigskip
\begin{acknowledgments}
\textbf{Acknowledgments.}
This work is supported by the Fermi Research Alliance, LLC under Contract No. DE-AC02-07CH11359 with the U.S. Department of Energy, Office of Science, Office of High Energy Physics. I.H. received support from the Brinson Foundation.
\end{acknowledgments}

% For when including just prebuilt bbl file for PRL or ArXiv
\input{LeptonDM-PRL-vfinal.bbl}
% For when building bilbiography
% \bibliography{LeptonDM-PRL-vfinal}

% Include supplemental material for arxiv
% \input{LeptonDM-PRL-vfinal-supplemental}

\setcounter{equation}{0}
\setcounter{figure}{0}
\renewcommand{\theequation}{S.\arabic{equation}}
\renewcommand{\thefigure}{S.\arabic{figure}}

\onecolumngrid
\appendix

\clearpage
\section{Supplemental Material: The Case of Scalar Dark Matter}

In the main body of this {\it Letter}, we focused on the case of DM in the form of a Dirac fermion. We can similarly consider a DM candidate that is a complex scalar, $\phi$, which couples to the same $L_{\mu}-L_{\tau}$ gauge boson considered above. In this case, the DM current is $J_{\rm DM}^\mu \equiv  i( \phi^* \partial^\mu \phi-\phi \partial^\mu \phi^*)$, and the annihilation cross section is
\begin{equation}
\sigma(s) = \sum_f
\frac{k_f {g^\prime}^4}{12\pi s}
%\sqrt{\frac{1-4m_f^2/s}{1-4m^2_{\chi}/s}}
\beta_f \, \beta_{\chi}
\left[\frac{
 (s + 2m_f^2)
}{ (s- m^2_{\zp})^2 + m_{\zp}^2 \Gamma^2_{\zp}} \right],~~~~
\end{equation}
where the sum is over SM species $f = \mu,\tau, \nu_{\mu,\tau}$ and the total $\zp$ width is
\begin{align}
\Gamma_{\zp} =  \frac{{g^\prime}^2 m_{\zp}}{48\pi} \bigg(1 - \frac{ 4m^2_{\phi}}{m^2_{Z'}}    \bigg)^{3/2}
+ \sum_f \Gamma_{\zp \rightarrow f\bar{f}} ,
\end{align}
where $\Gamma_{\zp \rightarrow f\bar{f}}$ is given in Eq.~\ref{gamma}. Unlike in the Dirac fermion scenario, here the annihilation cross section is $p$-wave, so annihilation during the CMB era is velocity suppressed. For scalar annihilation to $e^+e^-$ through kinetic mixing, the non-relativistic cross section satisfies
\begin{equation}
\sigma v_{\phi\phi^* \to e^+e^-} \simeq \frac{ (\varepsilon e g^\prime)^2 \, v^2  (m_e^2 + 2m_\phi^2 )
\sqrt{1 - m_e^2/m_{\phi}^2}
 }{  12 \pi \left[  (4m_\phi^2 - m_{\zp}^2)^2
+ m_{\zp}^2 \Gamma_{\zp}^2
   \right] },
\end{equation}
which is further suppressed by $v^2 \ll 1$ relative to the analogous expression for Dirac DM in Eq.~\ref{sigmav-cmb}, so this scenario is not constrained by CMB bounds on $p_{\rm ann}$.

\begin{figure}[t]
  \hspace{-0.3cm}
  \includegraphics[width=8.4cm]{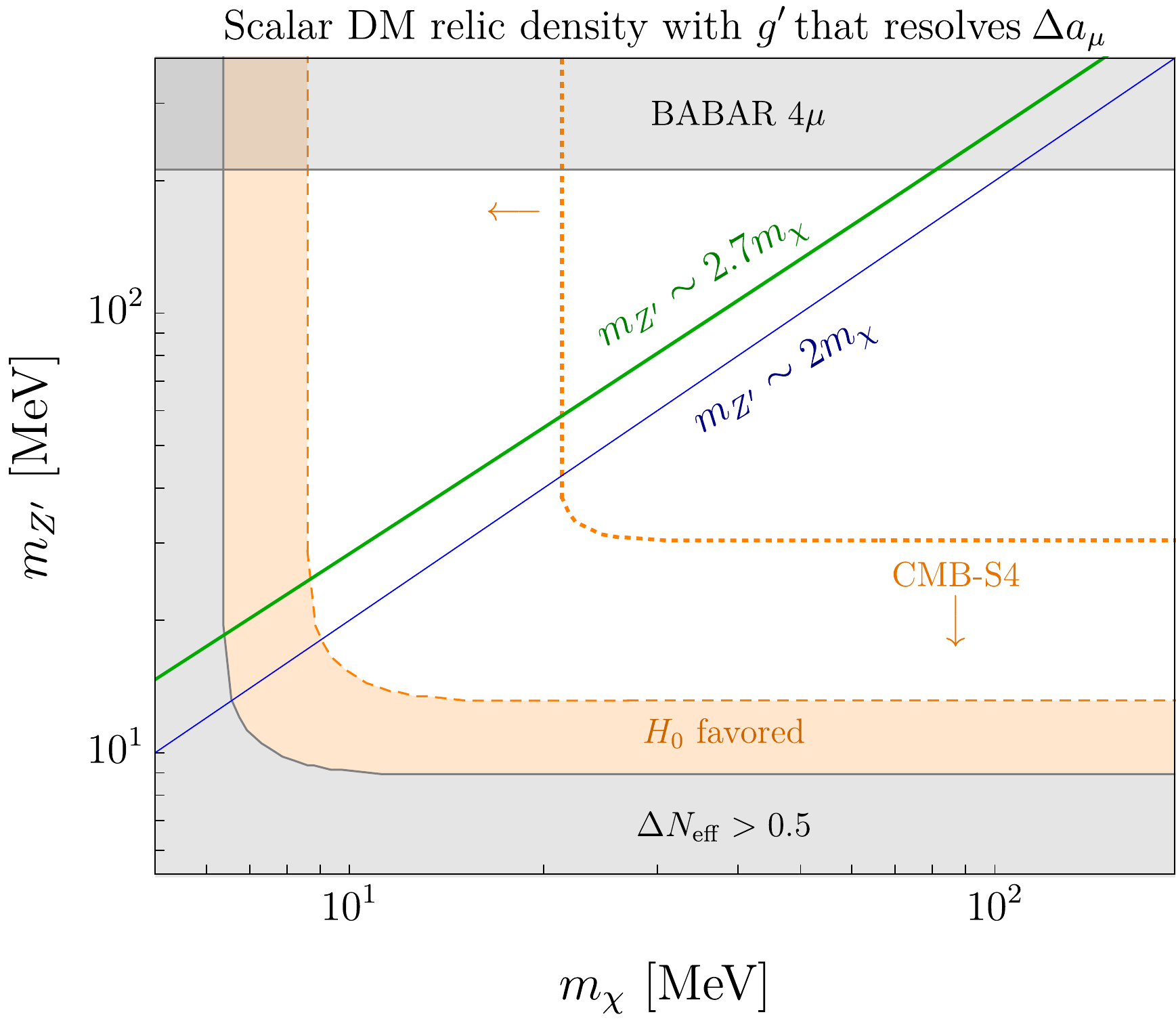}
  \caption{As in Fig.~\ref{fig:mainfig} of the main text, but for a complex scalar DM candidate instead of a Dirac fermion.}
  \label{sup-fig}
  \vspace{0cm}
\end{figure}

In Fig. \ref{sup-fig}, we show the parameter space for which complex scalar DM can achieve the observed DM relic density via thermal freeze out. These results are similar to those shown in Fig. \ref{fig:mainfig}, but the viable parameter space is slightly shifted relative to the Dirac fermion case due to the cross section being smaller. Here the constraints based on $\Delta N_{\rm eff}$ are the same as those described in the main text, except that the contribution from DM annihilation is rescaled by a factor of $1/2$ due to the difference in the number of spin degrees-of-freedom. The contribution to $\Delta N_{\rm eff}$ from $\zp$ decays is the same as described in the main text.

\clearpage
\section{Supplemental Material: Post-Freeze Out Annihilation Near Resonance}

\begin{figure}[t]
  \centering
  \includegraphics[width=0.6\linewidth]{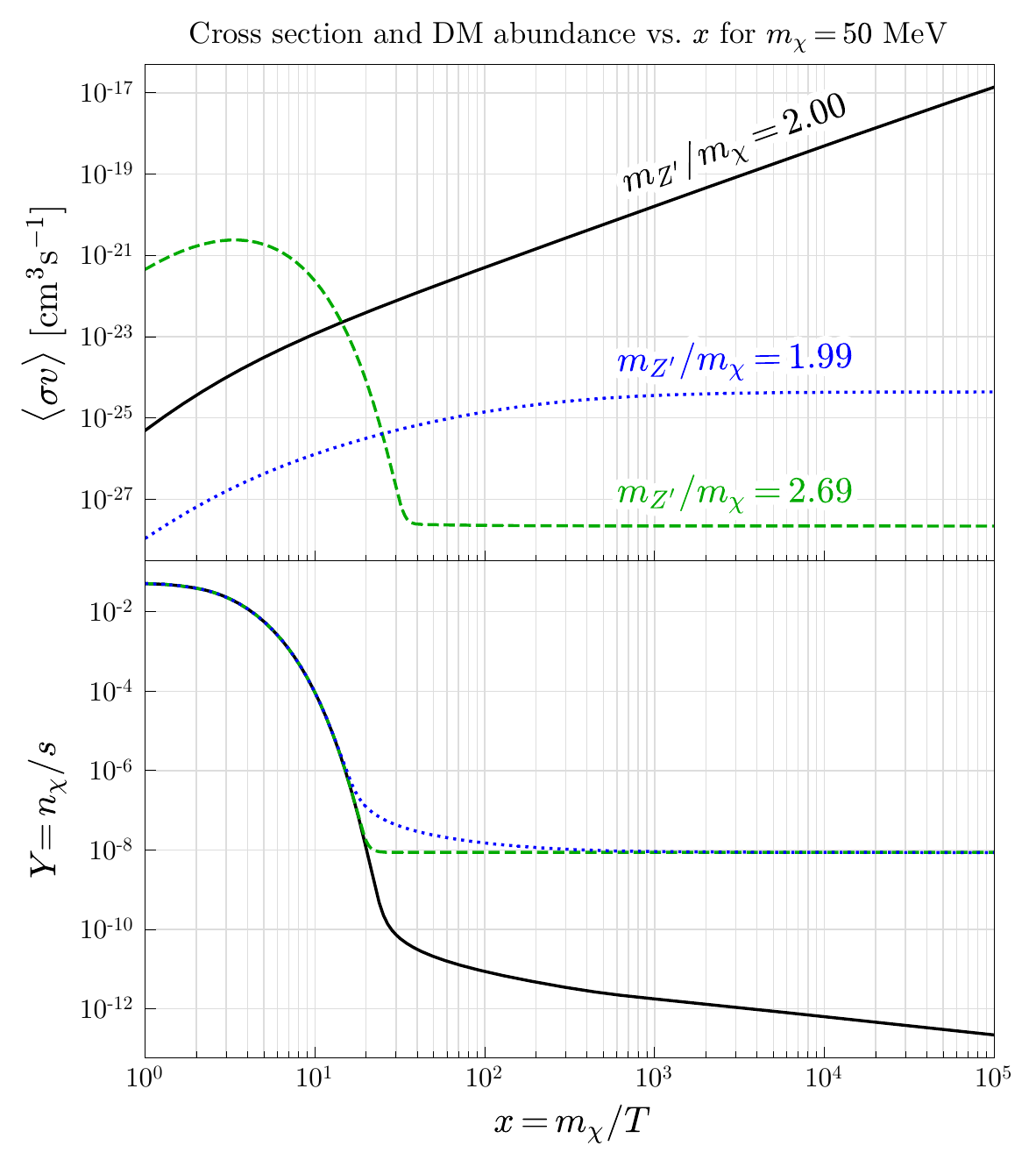}
  \caption{The top panel shows how the thermally averaged annihilation cross section evolves with the scaled inverse temperature, for three different mass ratios with $m_\chi = 50$ MeV. The bottom panel shows the evolution of the number density of dark matter per unit entropy from a numerical solution to the Boltzmann equation. The black (solid) line has $m_\zp = 2 m_\chi$, while the blue (dotted) and green (dashed) lines are the two mass ratios that produce the measured dark matter density in our universe for values of $g^\prime$ that resolve the muon $g-2$ anomaly. Note that
 even though DM freeze out occurs around $x \sim 20$, the number density is severely Boltzmann suppressed {\it well before}
 freeze out. This is important because for lighter DM masses near  $\sim$ 15 MeV, freeze out actually occurs after neutrino decoupling,
 but this is not catastrophic because the DM population is already exponentially suppressed by $T_{\rm dec} \sim 2$ MeV even
 though it's still in equilibrium.
    % Note that the black line shows significant annihilation still occurring long after freeze out.
  }
  \label{fig:resonantfreezeout}
\end{figure}

With an approximately constant annihilation cross section, a thermal relic abundance will typically settle to its final value soon after freeze out. However, when the $m_\zp/m_\chi$ mass ratio is very close to 2, resonant annihilation actually becomes more efficient as the temperature decreases, leading to a significant amount of post-freeze out annihilation. Fig.~\ref{fig:resonantfreezeout} illustrates this phenomenon in the solid black line, in contrast to the two other cases shown in green and blue.
% , which are sufficiently far from resonance during and after freeze out.
Note that even though the near-resonance cases result in the same final relic abundance, they take different amounts of time to reach a more stable value due to the increasing or decreasing behavior of $\langle \sigma v \rangle$ during freeze out.

In the low-velocity limit, the cross section has the following velocity dependence:
\begin{equation}
  \sigma v \propto \frac{1}{(4 - r^2 + v^2)^2 + r^4 \gamma_\zp^2},
\end{equation}
where $r = m_\zp / m_\chi$ and $\gamma_\zp = \Gamma_\zp/m_\zp$. For $r \le 2$, the cross section will increase as the velocity decreases until the two terms in the denominator are of similar magnitude, so the ultimate stopping point is controlled by the decay width. This occurs when $v \sim \sqrt{r^2 - 4 + r^2 \gamma_\zp}$ is suppressed in the thermal distribution.
% , which for a non-relativistic Maxwell-Boltzmann distribution, a velocity is suppressed when $x \gg \sqrt{12/v} \sim \sqrt{6/\sqrt{\gamma_\zp}}$.
Thus, post-freeze out annihilation will end earlier for $r < 2$ than for $r = 2$. For $r > 2$, the cross section will increase as the resonant peak becomes thermally favored, but then decrease and level out when typical velocities become too low to satisfy the resonance condition. All three cases in $r$ are illustrated in Fig.~\ref{fig:resonantfreezeout}. Due to this effect, one must be careful when solving the Boltzmann equation near resonance, since it may need to be integrated out to higher values of $x$ to find acceptable convergence.

\end{document}

%% file: LeptonDM-PRL-vfinal.bbl
%Control: production of eprint (1) enabled
%